\documentclass[prd,nofootinbib,floatfix,preprintnumbers]{revtex4}
\usepackage{epsfig}
\usepackage[dvips]{color}

\newcommand{\rcs}{\chi^2_{\rm red}}

\newcommand{\A}{\mathcal{A}}
\newcommand{\B}{\mathcal{B}}

\makeatletter
\renewcommand\@biblabel[1]{#1.}
\makeatother

\begin{document}
\title{A statistical mechanics approach to reverse engineering: sparsity and biological
priors on gene regulatory networks}
\author{M. Pica Ciamarra$^{1}$\footnote{Corresponding author - tel. +39 081 676805; fax +39 081 676346; picaciam@na.infn.it},
G. Miele$^{1,2}$\footnote{tel. +39 081 676463; fax +39 081 676463;
miele@na.infn.it}, L. Milano$^{1}$\footnote{tel. +39 081 676142;
milano@na.infn.it}, M. Nicodemi$^{1,3}$\footnote{tel. +39 081
676475; nicodem@na.infn.it}, G. Raiconi$^{4}$\footnote{tel. +39
089 963320; fax +39 089 963303; gianni@unisa.it}}

\affiliation{$^1$Universit\`{a} ``Federico II", Dipartimento di Scienze Fisiche, Napoli, Italy \& INFN Sezione di Napoli\\
$^2$Instituto de F\'{\i}sica Corpuscular (CSIC-Universitat de
Val\`{e}ncia), Ed.\ Institutos de Investigaci\'{o}n, Apdo.\ 22085,
E-46071 Val\`{e}ncia, Spain.\\
$^3$ Department of Physics and Complexity Science, University of Warwick, UK.\\
$^4$ Universit\'a di Salerno, Dipartimento di Matematica e
Informatica, Fisciano, Salerno, Italy}

\begin{abstract}
The important task of determining the connectivity of gene
networks, and at a more detailed level even the kind of
interaction existing between genes, can nowadays be tackled by
microarraylike technologies. Yet, there is still a large amount of
unknowns with respect to the amount of data provided by a single
microarray experiment, and therefore reliable gene network
retrieval procedures must integrate all of the available
biological knowledge, even if coming from different sources and of
different nature. In this paper we present a reverse engineering
algorithm able to reveal the underlying gene network by using
time-series dataset on gene expressions considering the system
response to different perturbations. The approach is able to
determine the sparsity of the gene network, and to take into
account possible {\it a priori} biological knowledge on it. The
validity of the reverse engineering approach is highlighted
through the deduction of the topology of several {\it simulated}
gene networks, where we also discuss how the performance of the
algorithm improves enlarging the amount of data or if any a priori
knowledge is considered. We also apply the algorithm to
experimental data on a nine gene network in {\it Escherichia
coli}. \end{abstract}

\maketitle \preprint{DSF/21/2007, IFIC-07-31}

\section{Introduction}

\noindent The amount and the timing of appearance of the
transcriptional product of a gene is mostly determined by
regulatory proteins through biochemical reactions that enhance or
block polymerase binding at the promoter region (\citealp{Jacob61,
Dickson75}). Considering that many genes code for regulatory
proteins that can activate or repress other genes, the emerging
picture is conveniently summarized as complex network where the
genes are the nodes, and a link between two genes is present if
they interact. The identification of these networks is becoming
one of the most relevant task of new large-scale genomic
technologies such as DNA microarrays, since gene networks can
provide a detailed understanding of the cell regulatory system,
can help unveiling the function of previously unknown genes and
developing pharmaceutical compounds.

Different approaches have been proposed to describe gene networks
(see (\citealp{filkov}) for a review), and different procedures
have been proposed
(\citealp{Tong02,Lee02,Ideker01,Davidson02,Arkin97,Yeung02}) to
determine the network from experimental data. This is a
computationally daunting task, which we address in the present
work. Here we describe the network via deterministic evolution
equations (\citealp{Tegner03, Bansal06}), which encode both the
strenght and the direction of interaction between two genes, and
we discuss a novel reverse engineering procedure to extract the
network from experimental data. This procedure, though remaining a
quantitative one, realizes one of the most important goal of
modern system biology, which is the integration of data of
different type and of knowledge obtained by different means.

We assume that the rate of synthesis of a transcript is determined
by the concentrations of every transcript in a cell and by
external perturbations. The level of gene transcripts is therefore
seen to form a dynamical system which in the most simple scenario
is described by the following set of ordinary differential
equations (\citealp{deJong02}):
\begin{equation}
\dot{X}(t) = \A X(t) + \B U(t)         \label{eq-cont}
\end{equation}
where $X(t) = (x_1(t),\ldots,x_{N_g}(t))$ is a vector encoding the
expression level of $N_g$ genes at times $t$, and $U$ a vector
encoding the strength of $N_p$ external perturbations (for
instance, every element $u_k$ could measure the density of a
specific substance administered to the system). In this scenario
the gene regulatory network is the matrix $\A$ (of dimension $N_g
\times N_g$), as the element $\A_{ij}$ measures the influence of
gene $j$ on gene $i$, with a positive $\A_{ij}$ indicating
activation, a negative one indicating repression, and a zero
indicating no interaction.

The matrix $\B$ (of dimension $N_g \times N_p$) encodes the
coupling of the gene network with the $N_p$ external
perturbations, as $\B_{ik}$ measures the influence of the $k$-th
perturbation on the $i$-th gene.

A critical step in our construction is the choice of a linear
differential system. Even if a such kind of model is based on
particular assumptions on the complex dynamics of a gene network,
it seem the only practical approach due to the lack of knowledge
of real interaction mechanism between thousands of genes. Even a
simple nonlinear approach would give rise to an intractable amount
of free parameters. However, it must also be recognized that all
other approaches or models have weakness points. For instance,
boolean models (which have been very recently applied to inference
of networks from time series data, as in (\citealp{martin}),
strongly discretize the data and select, {\it via} the use of an
arbitrary threshold, among active and inactive gene at every
time-step. Dynamical Bayesian models, instead, are more data
demanding than linear models due to their probabilistic nature.
Moreover, their space complexity grows like $N_g^4$ (at least in
the famous Reveal Algorithm by K.P. Murphy (\citealp{Murphy01})),
which makes this tool suitable for small networks.

The linear model of Eq. (\ref{eq-cont}) is suitable to describe
the response of a system to small external perturbations. It can
be recovered by expanding to first order, and around the
equilibrium condition $\dot{X}(t) = 0$, the dependency of
$\dot{X}$ on $X$ and $U$, $\dot{X}(t) = f(X(t),U)$. Stability
considerations ($X(t)$ must not diverge in time) require the
eigenvalues of $\A$ to have a negative real part. Moreover it
clarifies that if the perturbation $U$ is kept constant the model
is not suitable to describe periodic systems, like cell cycles for
example, since in this case $X(t)$ asymptotically approaches a
constant.

Unfortunately data from a given cell type involve thousands of
responsive genes $N_g$. This means that there are many different
regulatory networks activated at the same time by the
perturbations, and the number of measurements (microarray
hybridizations) in typical experiments is much smaller than $N_g$.
Consequently, inference methods can be successful, but only if
restricted to a subset of the genes (i.e. a specific network)
(\citealp{Basso05}), or to the dynamics of genes subsets. These
subsets could be either gene clusters, created by grouping genes
sharing similar time behavior, or the modes obtained by using
singular value decomposition (SVD). In these cases it is still
possible to use Eq. (\ref{eq-cont}), but $X(t)$ must be
interpreted as a vector encoding the time variation of the
clusters centroids, or the time variation of the characteristics
modes obtained via SVD.

In this paper we present a method for the determination of the
matrices $\A$ and $\B$ starting from time series experiments using
a Global Optimization approach to minimize an appropriate figure
of merit.  With respects to previous attempts, our algorithm as
the uses explicitly the insight provided by earlier studies on
gene regulatory networks (\citealp{Barabasi00,Barabasi01}),
namely, that gene networks in most biological systems are sparse.
In order to code such type of features the problem itself must be
formulated as mixed-integer nonlinear optimization one
(\citealp{minlp}). Moreover our approach is intended to explicitly
incorporate prior biological knowledge as, for instance, it is
possible to impose that:  $\A_{ij} < 0$ $(= 0,> 0, \neq 0)$ if it
is known that gene $j$ inhibits (does not influence, activates,
influences) gene $i$. This means that the optimization problem is
subject to inequality and/or equality constraints. Summing up the
characteristics of the problem we must solve: high dimensionality,
mixed integer, nonlinear programming problem for the exact
solution of which no method exists. An approximate solution can be
found efficiently using a global optimization techniques
(\citealp{Pardalos95, Pardalos02}) based on an intelligent
stochastic search of the admissible set. As consequence of the
optimization method used, there is no difficulties to integrates
different time series data investigating the response of the same
set of genes to different perturbations, even if different time
series are sampled at different (and not equally spaced) time
points. The integration of different time series is a major
achievement, as it allows for the joint use of data obtained by
different research groups.We believe that the integration of
multiple time-series dataset in unveiling a gene network is a
topic of great interest as focused in recently published papers
(\citealp{shi}).

We illustrate and test the validity of our algorithm on computer
simulated gene expression data, and we apply it to an experimental
gene expression data set obtained by perturbing the SOS system in
bacteria {\it E. coli}.

\section{Methods}

The simplest assumption regarding the dynamical response of gene
transcripts (intially in a steady state, $X(t) = 0$ for $t < 0$),
to the appearance of an external perturbation $U(t)$ at time $t>0$
is given by Eq. (\ref{eq-cont}). Since the state of the system
measured at discrete times $t = t_k$, $k=0,\ldots,N_t$, it useful
to consider the discrete form of Eq.~\ref{eq-cont}.
\begin{equation}
\label{eq-discrete} X(t_{k+1}) = A X(t_k) +
\widetilde{U}(t_k,t_{k+1}),
\end{equation}
where $A$ is a matrix with dimension $N_g \times N_g$,
and $\widetilde{U}$ is a function of the perturbations, namely

\begin{eqnarray}
A &=& \exp(\A \Delta t),\nonumber\\
\widetilde{U}(t_k,t_{k+1}) &=&
\int_{t_k}^{t_{k+1}}\exp\{\A(t_{k+1}- \tau)\} \, \B \, U(\tau) \,
d\tau .\label{eq-discrete-matrix}
\end{eqnarray}
Here we have assumed, for simplicity sake, $t_k = k\Delta$, but
the generalization to the most general case is straightforward. In
particular, for constant $U$ one gets $B \equiv
\widetilde{U}(t_k,t_{k+1})=\left( \exp\{ \A \Delta t\}-1 \right)
\A^{-1} \B \, U$.

Due to the presence of noise the measured $X(t_k)$ do not coincide
with the true values $\overline{X}(t_k)$ expected to satisfy Eq.
(\ref{eq-cont}). If we for simplicity observed samples affected by
 independent, zero mean additive noise $\varepsilon_{k}$, namely $X(t_k) =
\overline{X}(t_k)+\varepsilon_{k}$, the matrices ruling the
dynamics of Eq. (\ref{eq-cont}) can be found by requiring the
minimization of a suitably defined {\it cost function}.

Under the simplifying assumption of a constant external
perturbation, previous works have been focused on the
determination of $A$ and $B$ (from which $\A$ and $\B$ can be
retrieved), as in (\citealp{Bansal06,Holter}). The matrices $A$
and $B$ have been assumed to be those minimizing the cost function
\begin{equation}
CF(A,B) = \sum_{k = 0}^{N_t-1} |X(t_{k+1})-(AX(t_k)+BU))|^2.
\label{eq-cost}
\end{equation}
Eq.~\ref{eq-cont} can be written as standard linear least squares
estimation problem for $A, B$, whose solution can be found by
computing the pseudoinverse of a suitable matrix, providing that
number of observations is sufficiently high: $N_{t}>N_{g}+N_{p}$.

In the present analysis we introduce a new reverse engineering
approach to determine the matrices $\A$ and $\B$, which turns out
to be more efficient and flexible than previous ones. Our approach
is based on the following considerations, which have not been
taken into account in previous works:
\begin{itemize}
\item[a.] Each gene expression time--series could in principle be scanned according
to both time versus, namely the {\it time-reversibility} of
dynamics.
\item[b.] There is a biological evidence suggesting that the matrix $A$ is sparse (\citealp{Barabasi00,Barabasi01}).
For this reason any reverse engineering algorithm has to be able
to capture the proper sparsity of gene regulatory network.
\item[c.] In many situations there are biological prior information about
bounds on numerical value of some specific entries  of $\A$ and
$\B$. Such bounds must be taken into account in the solution
procedure.
\end{itemize}
As a consequence of this we use as a cost function the reduced $\rcs$ defined
as follows:
\begin{equation}
\label{eq-rcs}
\rcs = \frac{\mathcal{CF}(\A,\B)}{n_{\rm dof}\sigma^2},
\end{equation}
where
\begin{eqnarray}
\mathcal{CF}(\A,\B) &=&\sum_{k = 0}^{N_t-1}\left[
\left|X(t_{k+1})-\left(A \,
X(t_k)-\widetilde{U}(t_k,t_{k+1})\right) \right|^2 +
\left|X(t_k)-\left(A^{-1}X(t_{k+1}) +
\widetilde{U}(t_{k+1},t_k)\right)\right|^2\right].
\label{eq-mcost}
\end{eqnarray}
Note that $A^{-1}\widetilde{U}(t_k,t_{k+1}) =
-\widetilde{U}(t_{k+1},t_k)$, and the quantities $A$ and
$\widetilde{U}(t_k,t_{k+1})$ can be obtained from $\A$ and $\B$ by
appropriate numerical approximation algorithms for
Eq.s~(\ref{eq-discrete-matrix}). The quantity $\sigma$ denotes the
standard deviation of the independent, additive noise affecting
the dataset.

A straightforward optimization on dynamics/input matrices of Eq.
(\ref{eq-cont}) is the main improvement of the proposed approach
with respect to the previous ones. This is the only way that
enable us to incorporate the sparseness requirement on $\A,\B$ and
eventually available biological priors. It is clear that
sparseness was destroyed by exponentiation and integration
involved in the continuous-discrete transformation of the problem,
in the same way simple bounds on $\A,\B$ elements are transformed
in highly complex nonlinear relations on $A,B$. The price paid for
the flexibility of the approach is the computational effort
required for any computation of the error function. This put a
very strong attention to the efficiency of the optimization
algorithm.

In Eq.~(\ref{eq-mcost}) the two contributions in square brackets
account for the forward and backward propagation, respectively,
and thus implement the time reversibility of the dynamics.
Moreover, the sparsity of the gene network is taken into account
via the number of degrees of freedom (d.o.f.) defined as $n_{\rm
dof} \equiv n_{\rm par} - n_{\rm eq}$ with $n_{\rm par} = N_g(N_g
+ N_p) - n_{\rm zero}$ the number of free parameters, $n_{\rm eq}
= N_g  (N_t - 1)$ the number of equations (constraints) and
$n_{\rm zero}$ the number of elements of $\A$ and $\B$ (a total of
$n_{\rm zero}$) fixed to zero.

The generalization of the algorithm to the case in which there are different
time-series, $X^\alpha(t_k)$, corresponding to the response of the same set of
genes to similar and/or different perturbations $B^\alpha$ with $\alpha = 1, . . . ,N_p$ is
straightforward. In this case the cost function to be minimized is simply
\begin{equation}
 \rcs = \frac{1}{2 n_{\rm dof}} \sum_{\alpha = 1}^{N_p}\frac{\mathcal{CF}(\A,\B^\alpha)}{\sigma^2_\alpha}.
\label{eq-min-molti}
\end{equation}
Here we have assumed the noise to depend on the time-series ($\alpha$). It is clearly
possible, however, to introduce a time ($t_k$) and even a gene (i) dependence, i.e to use $\sigma = \sigma^i_\alpha(t_k)$.

We detail now our procedure to find the spare matrices $\A$ and
$\B$ minimizing $\rcs$,, which is in general a formidable task.
The first difficulty is the determination of the number $n_{\rm
par}$ of not vanishing elements of $\A, \B$ (or equivalently the
number of d.o.f. $n_{\rm dof}$ ). Having determined $n_{\rm par}$
the problem is still very complicated since there are
\begin{equation}
\frac{(N_g(N_g+N_p))!}{n_{\rm par}! \, (N_g(N_g+N_p)-n_{\rm
par})!}
\end{equation}
different ways of choosing these $n_{\rm par}$ elements out of the
$N_g(N_g+N_p)$ candidates. For typical values of the parameters, for
instance $N_g = 10$ and $n_{\rm par} = 1/2N_g^2 = 50$, the number of
possible combinations is of the order of $10^{32}$, so big that any
kind of extensive algorithmic procedure is precluded.
 A practical approach to, at least approximately solve, this formidable problem
is that of resort to  a global optimization techniques based on a
stochastic strategy to search of the admissible set, for a
comprehensive review os such type of methods one can see
(\citealp{Pardalos95,Pardalos95}). We have tackled this problem
via the implementation of the more classical of such methods:  a
simulated annealing procedure (\citealp{Kirkpatrick}), based on a
Monte Carlo dynamics. For each possible value of the number of
parameters $n_{\rm par}$, the algorithm search for the matrices
$\A$ and $\B$ with a total of $n_{\rm par}$ non zero elements
minimizing the cost function of Eq. (\ref{eq-mcost}), as discussed
below. We then easily determine $\rcs(n_{\rm par})$ and the
minimizing matrices $\A^*$ and $\B^*$ which are our best estimates
of the true matrices. In order to determine the matrices $\A$ and
$\B$ with a total of $n_{\rm par}$ non zero parameters which
minimize the cost function, our simulated annealing procedure
starts with two random matrices $\A$ and $\B$ with a total of
$n_{\rm par}$ not vanishing parameters, and changes the elements
of these matrices according to two possible Monte Carlo moves. One
move is the variation of the value of a not vanishing element of
the two matrices, the other one consists in setting to zero a
previously non-zero element, and to a random value a zero element.
Each move, which involves a variation $\Delta \mathcal{CF}$ of the
cost function, is accepted with a probability $\exp[-\Delta
\mathcal{CF}/T]$, where $T$ is an external parameter. As in
standard optimization by annealing procedures, we start from a
high value of $T$, of the order of the cost function value, and
then we slowly consider the limit $T \to 0$. In the limit of
infinitesimally small decrease of $T$ the algorithm is able to
retrieve the true minimum of the cost function, while for faster
cooling rates estimates of the real minimum are recovered.

As the Monte Carlo moves attempt to change the values of the
elements of $\A$ and $\B$, it is easy to introduce biological
constraints on the values of $A_{ij}$ and of $B_k$, as we will shown
in a following example. The algorithm requires the evaluation of the
cost function $\mathcal{CF}$, which is a time consuming operation as
the computation of the discrete matrix $A$ and of its inverse
$A^{-1}$ are required. We have implemented this algorithm in C++
making use of the GNU Scientific Library, www.gsl.org.

\section{Results}
In this section, we illustrate our reverse engineering algorithm
with three examples. The validity of our algorithm and of other
known ones are evaluated by comparing the exact dynamical matrices
$\A$ and $\B$ with their best estimate $\A^*$ and $\B^*$ obtained
via the reverse engineering procedure. To this end, we have
introduced the parameter
\begin{equation}
\eta_{\mathcal{C}} =
\frac{|\mathcal{C}-\mathcal{C}^*|}{|\mathcal{C}|},
\label{eq-parameter}
\end{equation}
where $\mathcal{C}^* = \A^*$ or $\B^*$ and $|\mathcal{C}|$ is the
$L_2$ norm of the matrix $\mathcal{C}$. Clearly, $\eta_C \ge 0$,
the equality being satisfied if and only if $C = C^*$. Since
$\eta_C$ is a measure of a relative error it has no upper bound,
but the estimate of $C$ becomes unreliable when $\eta_C$ is above
$1$, i.e. $|C - C^*| > |C|$. This parameter allows for a faithful
evaluation of the quality of the reverse engineering approach, as
it summarizes the comparisons of all retrieved elements $C_{ij}$
with their true values $C^*_{ij}$.

We discuss three applications. First, we show how our algorithm
works when applied to a single time series. In this case one can
show that the cost function $\rcs(\A,\ B)$, which takes into
account both the forward and the backward propagation, is more
effective in determining the structure of the gene network than
the usual cost function $CF(A,B)$ of Eq. (\ref{eq-cost}), which
only considers the forward propagation. The second example shows
how we can easily take into account the presence of different
time-series, while the last example shows how biological priors
can be included. Before discussing the examples we shortly
describe the procedure used to generate the synthetic dataset.

\subsubsection{Generation of a synthetic dataset} In order to
generate a synthetic dataset $X(t_k)$ one must construct the
matrices $\A$ and $\B$, from which it is possible to generate the
noiseless time-series $\overline{X}(t_k)$. Hence, one gets $X(k) =
\overline{X}(t_k) + \varepsilon_k$ for $k = 1, ..,N_t$ where
$\varepsilon_k$ are i.i.d. random variables with standard
deviation $\sigma$.

While there are no constraints on $\B$, $\A$ must be a sparse
random matrix whose complex eigenvalues have negative real part.
The generation of $\A$ proceeds according the following steps.
First, we generate a $N_g \times N_g$ block diagonal matrix
$\A^{(0)}$, whose $N_g$ blocks are $2 \times 2$ antisymmetric
matrices with diagonal elements $\lambda_r^\alpha$ and off
diagonal elements $\lambda_i\alpha$, or $1 \times 1$ negative real
elements $\lambda$. By direct constructions all of the $N_g$
eigenvalues of the matrix $\A^{(0)}$ have negative real part. Then
we generate a series $R_k$ of random unitary matrices, with only
$4$ off-diagonal not vanishing entries, and compute the matrices
$A^k = R_k\A^{(k-1)}R^{-1}_k$, all of them sharing the spectrum of
$A^{(0)}$. Clearly, as $k$ grows, the number of vanishing entries
(the sparsity) of $A^{(k)}$ decreases. We fix $\A$ as the matrix
$\A^{(k)}$ characterized by the desired number of vanishing
elements. By choosing typical values of $\lambda_r^\alpha$ and
$\lambda_i^\alpha$ it is possible to control the time scale of the
relaxation process of the system following the application of the
perturbation.
\begin{figure}[t]
\begin{center}
\includegraphics*[scale=1]{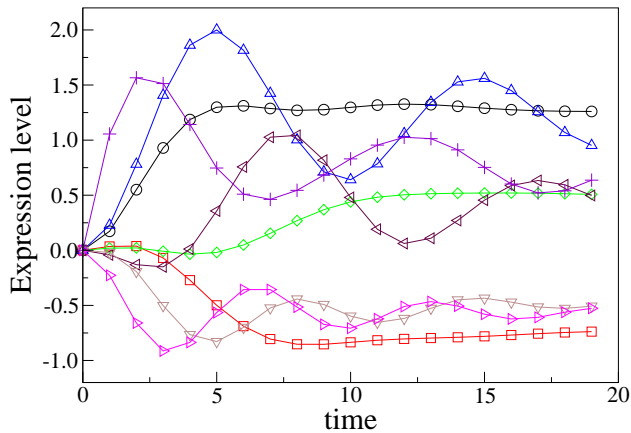}
\end{center}
\caption{\label{fig1} Synthetic time-series $X(t_k)$ with $N_g = 8$ elements measured at
$N_t$ = 20 equally spaced time-points.}
\end{figure}

\subsubsection{Example 1: a single time series}
Let us consider a simulated time-series $X(t_k)=
(x_1(t_k),\ldots,x_{N_g}(t_k))$ with $N_g = 8$ measured at $N_t =
20$ equally-spaced time-points, as shown in Fig.~\ref{fig1}. This
dataset is generated by starting from a sparse gene network $\A$
(with only $49$ out of $N_g^2 = 64$ non-zero elements), a constant
perturbation $U(t) = 1$ and a sparse external
perturbation-coupling matrix $\B$ with a single not vanishing
entry. The white noise is characterized by a standard deviation
\begin{equation}
\sigma(p) = p \sum_{i=1}^{N_g} \sum_{k=1}^{N_t} \frac{|
x_i(t_k)|}{N_g \, N_t}, \label{eq-noise}
\end{equation}
measured in units of the mean absolute value of the expression
levels of all genes. In particular the value $p = 0.05$ has been
used.

\begin{figure}[t]
\begin{center}
\includegraphics*[scale=1]{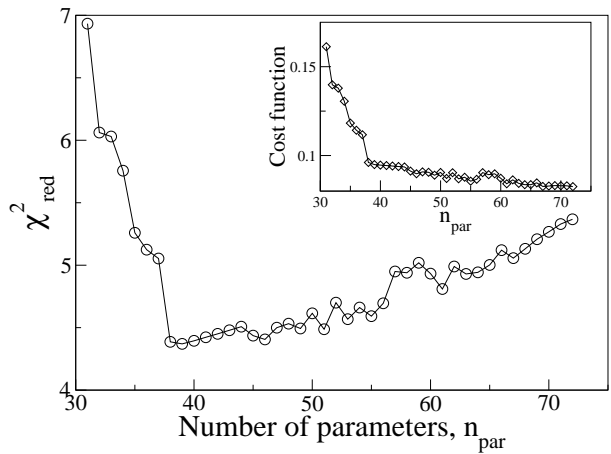}
\end{center}
\caption{\label{fig2} The main panel (inset) show the dependence
of $\rcs$ (of the minimum of the cost function) on the number of
not vanishing parameters $n_{\rm par}$, as determined by our
algorithm when applied to the time-series shown in
Fig.~\ref{fig1}. The fluctuations are due to the probabilistic
nature of the Monte Carlo minimization procedure. The quantity
$\rcs$ varies non-monotonically with $n_{\rm par}$, and has a
minimum with $n_{\rm par} = 39$ parameters.}
\end{figure}

We have applied our algorithm to this dataset. To this end, we
have minimized the reduced chi-square $\rcs$, defined in Eq.
(\ref{eq-rcs}), for different values of the number of parameters
$n_{\rm par}$ (i.e. of the number of degrees of freedom $n_{\rm
dof}$). Fig.~\ref{fig2} shows that $\rcs$ has a non-monotonic
dependence on the number of parameters $n_{\rm par}$. This feature
is a signature of the fact that both networks with few or with
many connections are bad descriptions of the actual gene
regulatory system. Accordingly, our best estimate of the number of
not vanishing parameters is $n_{\rm par}^* = 39$, where $\rcs$ has
its minimum, and the corresponding minimizing matrices $\A^*$
(with $33$ non zero entries) and $\B^*$ (with $6$ non zero
elements) are our best estimates of the actual gene network
encoding matrix $\A$ and of the matrix $\B$.
The estimators assume the values $\eta_\A = 0.76$ and $\eta_\B =
0.005$. These values indicate that, when applied to this small
dataset, our algorithm is able to retrieve $\B$ to a very good
approximation, and $\A$ with a comparatively larger error.

\begin{figure}[t]
\begin{center}
\includegraphics*[scale=1]{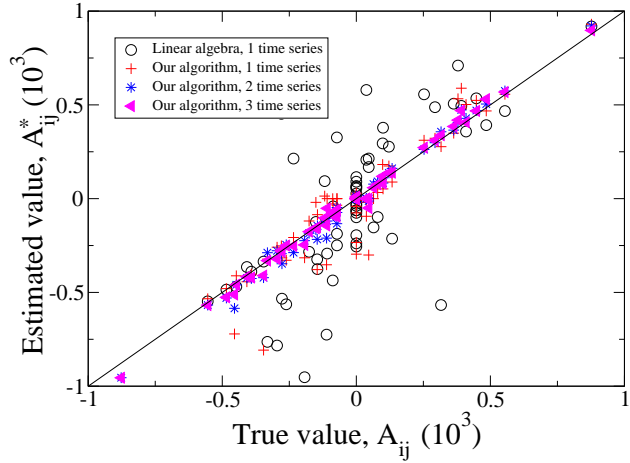}
\end{center}
\caption{\label{fig3} We plot here the values of the element of
the estimated matrices $A^*_{ij}$, obtained both with the linear
algebraic approach and with our algorithm, versus their true value
$A_{ij}$. Ideally, the points should line on the $y=x$ dotted
line.}
\end{figure}

For comparison, we have also obtained the matrices $A$ and $B$
which exactly minimize $CF(A,B)$ via a linear algebraic approach,
and retrieved the corresponding continuous matrices via the use of
the bilinear transformation, obtaining the scores $\eta_\A = 2.1$
and $\eta_\B = 0.012$. These numbers prove that by exploiting the
time reversibility of the equation of motion, and the sparseness
of the gene network, is it possible to estimate the parameters of
the network with a greater accuracy, as also shown in
Fig.~\ref{fig3} where we plot the best estimates $\A^*_{ij}$
obtained by both methods versus their true values $\A_{ij}$:
in the case of perfect retrieval all of the points should lie on the $y=x$ line.\\

\subsubsection{Example 2: multiple time series}
There are two major problems encountered when trying to infer a
gene network via the analysis of time-series data. The first one
is that there are usually to few time-points with respect to the
large number of genes. The second one is associated to the fact
that, when the system responds to an external perturbation, only
the expression of the genes directly or indirectly linked to that
perturbation changes, i.e., only a specific sub-network of the
whole gene network is {\it activated} by the external
perturbation. While through the study of the time-series it is
possible to learn something about the regulatory role of the
responding genes, nothing can be learnt about the regulatory role
of the non-responding genes.

These problems can be addressed by using gene network retrieval
procedures which are able to simultaneously analyze different
time-series (\citealp{Wang}), particularly if these measure the
response of the system to different perturbations, as we expect
different perturbations to activate different genes. Our reverse
engineering approach naturally exploits the presence of multiple
time series by requiring the minimization of Eq.
(\ref{eq-min-molti}).

Here we study the network discussed in the previous example by
adding to the time-series shown in Fig.~\ref{fig1}, other ones
generated by the application of two different perturbations. For
sake of simplicity all time-series are measured at equally-spaced
time-points, but with an elapsing time between two consecutive
data points depending on the particular time-series. Hence that
the problem cannot be reduced to the one of a single average
time-series by exploiting the linearity of Eq. (\ref{eq-cont}).

As the number of time-series increases, our determination of the
gene network $\A$ becomes more and more accurate. For instance,
while by means of a single perturbation we obtain $\eta_\A = 0.76$
($\eta_{\B_0} = 0.005$), by using two time-series we obtain
$\eta_\A = 0.25$ ($\eta_{\B_0} = 0.004, \eta_{\B_1} = 0.003$), and
by using three time series we get $\eta_\A = 0.13$ ($\eta_{\B_0} =
0.004, \eta_{\B_1} = 0.002, \eta_{\B_2} = 0.002$).
\begin{figure}[t]
\begin{center}
\includegraphics*[scale=1]{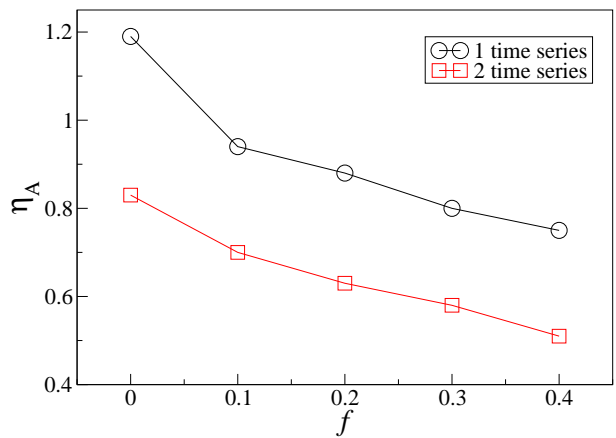}
\end{center}
\caption{\label{fig_priors} Dependence of $\eta_\A$ on the
fraction of priors, as obtained by analyzing one or two
time-series. The scoring parameter $\eta_\A$ decreases as the
number of priors increases, indicating that a better estimate of
the gene network $\A$ is recovered.}
\end{figure}\\
\\
\subsubsection{Example 3: biological priors}
As the traditional approach to research in Molecular Biology has
been an inherently local one, examining and collecting data on a
single gene or a few genes, there are now couples of gene which
are known to interact in a specific way, or do not interact at
all. This information is nowadays easily available by consulting
pubic databases such as Gene Ontology. Here we show that it is
possible to integrate this non-analytical information in our
reverse engineering approach, improving the accuracy of the
retrieved network. To this end we consider again the gene network
$\A$ but we introduce some constraints on a fraction $f$ of
randomly selected elements of the matrices $\A$ and $\B$, namely
$10\% \leq f \leq 40\%$. As our retrieval procedure tries to
exchange vanishing and not vanishing elements of $\A$ and $\B$ we
introduce the constraints as follows: if the element is zero in
the exact matrices then we set it to zero and we never try to set
it to a non-zero value; on the contrary, if the element is
different from zero, its value is free to change and we never try
to set it to zero. By using this approach we assure that our best
estimates of $\A$ and $\B$ are consistent with the previous
knowledge. In order to stress the greater improvement that can be
obtained via the use of biological priors, we consider now the
same gene network  $\A$ and perturbations of examples 1 and 2, but
we corrupt the noiseless dataset by adding a noise (see Eq.
(\ref{eq-noise})) characterized by $p = 0.1$, and not by $p =
0.05$ as before. Due to the high value of the noise the linear
algebraic approach is not more able to recover the gene network
matrix, as it obtains a score $\eta_\A = 4.40$.

We show in Fig.~\ref{fig_priors} the dependence of $\eta_\A$ on
the fraction of randomly selected elements of $\A$ and $\B$ fixed
either to zero or to non-zero, both for the case in which only one
or two perturbations have been used in the retrieval procedure. As
expected, $\eta_A$ decreases as the number of priors increases,
showing that as more biological knowledge on the system of
interest is available the reliability of our reverse engineering
approach improves.

\subsubsection{Results on Escherichia Coli}
\noindent We applied our algorithm to a nine gene network, part of
the SOS network in {\it E. Coli}. The genes are $recA$, $lexA$,
$Ssb$, $recF$, $dinI$, $umuDC$, $rpoD$, $rpoH$, $rpoS$, and the
used time-series consists of six time measurements (in triplicate)
of the expression level of these genes following treatment with
Norfloxacin, a known antibiotic that acts by damaging the DNA. The
time series is the same used in Ref. (\citealp{Bansal06}), and
experimental details can be found there.

Given $N_g = 9$ there are $90$ unknowns to be determined, as $\A$
is a $N_g \times N_g$ matrix, and $\B$ is a vector of length
$N_g$. Since $N_t = 6$, the experimental data allows for the
writing of $N_g(N_t -1) = 45$ equations, and for the determination
of only $45$ unknowns, while a literature survey
(\citealp{Bansal06}) suggests that there are at list $52$
connections between the considered genes (including the
self-feedback). As in previous works, we are therefore forced to
use an interpolation technique to add new time measurements,
creating a time series with $11$ time points.

When applied to this dataset, our algorithm found that $\rcs$ is
minimized by a matrix $\A$ with $57$ not vanishing entries, and a
vector $\B$ with $6$ non-zero elements, which are given in Table
\ref{table1}. In the literature, there are $52$ known connections
between the nine considered genes, including the self-feedback..
We are able to find $37$ of these connections. Regarding the
interaction with Norfloxacin, our algorithm found that primary
target is $recA$, as expected.
\begin{table*}[h!]
\begin{center}
\begin{tabular}{l|ccccccccc||c}
        & recA  & lexA  & Ssb   & recF  & dinI  & umuDC & rpoD  & rpoH  & rpoS & $\B$\\ \\
\hline
recA  & -1.68 & - & -0.36 & 1.81 & 1.05 & 0.84 & - & - & -0.59 & 0.71\\ \\
lexA  & -0.11 & -1.56 & 0.59 & 0.58 & 0.40 & - & -0.34 & - & - & 0.13\\ \\
Ssb  & -0.47 & 1.82 & -2.83 & - & 0.60 & - & 0.96 & -1.71 & 1.29 & -\\ \\
recF  & 0.68 & 0.42 & - & -0.93 & -0.52 & -0.40 & -0.30 & 1.13 & - & 0.38\\ \\
dinI  & 1.18 & 0.72 & 0.39 & -0.96 & -1.71 & 0.42 & - & - & - & 0.34\\ \\
umuDC  & 0.47 & -0.63 & -0.39 & -0.64 & 0.19 & -0.65 & 0.11 & - & 0.53 & -\\ \\
rpoD  & -0.06 & -0.28 & - & 0.36 & - & - & -0.22 & - & - & 0.40\\ \\
rpoH  & - & - & -1.10 & 1.60 & -0.32 & 0.92 & - & -3.46 & 1.46 & -\\ \\
rpoS  & -0.39 & -0.43 & - & - & 0.18 & 0.92 & 0.26 & 0.82 & -0.72 & -0.11\\
\end{tabular}
\end{center}
\caption{The matrix $\A$ encoding the SOS network for {\it E.
coli}: each element codes the effect of the gene of the column on
the gene of the raw. The last column shows the effect of
Norfloxacin on the considered genes, $\B$. All elements are
expressed in $10^{-2} s^{-1}$. The matrix $\A$ has $57$ not
vanishing elements, while $\B$ has $6$ non-zero elements. For
visualization purposes zero elements have been replaced by a dash
`-'. } \label{table1}
\end{table*}

\section{Conclusions}
In the framework of a linear deterministic description of the time
evolution of gene expression levels, we have presented a reverse
engineering approach for the determination of gene networks. This
approach, based on the analysis of one or more time-series data,
exploits the time-reversibility of the equation of motion of the
system, the sparsity of the gene network and previous biological
knowledge about the existence/absence of connections between
genes. By taking into account this information the algorithm
significatively improves the level of confidence in the
determination of the gene network over previous works.

The drawback of our procedure is the computational cost, which at
the moment limits the applicability of the algorithm to a small
number of genes/clusters. There are two time-consuming procedures.
One is the transformation of the continuous matrix $\A$ in the
discrete matrix $A$, which we have been avoided by using the
bilinear transformation, but whose validity breaks down as the
time interval between two consecutive measurements increases. The
second one, which at the moment is the most expensive in time, is
the computation of the inverse matrix $A^{-1}$, which we
accomplish through the so-called LU decomposition whose
computational cost is $O(N^3)$. Alternative methods for exploiting
the reversibility of the dynamics should therefore by devised for
applications with a larger number of genes.\\

\section*{Acknowledgments}
\noindent We thank D. di Bernardo for rousing our interest in this
subject, and for helpful discussions.

\end{document}